\documentclass[10pt]{raa}            
\usepackage[T1]{fontenc}
\usepackage{ae,aecompl}
\usepackage{graphicx,times}             
\usepackage{amsmath}
\usepackage{cases}
\begin{document}

   \title{Investigating the morphology of the supernova remnant G349.7+00.2 in the medium with a density gradient
}

   \volnopage{Vol.0 (20xx) No.0, 000--000}      
   \setcounter{page}{1}          

   \author{Jing-Wen\ Yan
      \inst{1}
   \and Chun-Yan\ Lu
      \inst{1}
   \and Lu\ Wen
      \inst{1}
   \and Huan\ Yu
      \inst{2}
   \and Jun\ Fang
      \inst{1}
   }

   \institute{Department of Astronomy, Key Laboratory of Astroparticle Physics of Yunnan Province, Yunnan University, Kunming 650091, China; {\it fangjun@ynu.edu.cn}\\
        \and
             Department of Physical Science and Technology, Kunming University, Kunming 650214, China; {\it yuhuan.0723@163.com}\\
   }

   \date{xxx}

\abstract{
G349.7 + 00.2 is a young Galactic supernova remnant (SNR) with a mushroom morphology in radio and X-rays, and it has been detected across the entire electromagnetic spectrum from radio to high energy $\gamma$-rays. Moreover,  the remnant is interacting with a molecular cloud based on the observations in the radio and infrared band. The reason for the formation of the periphery and the dynamical evolution of the remnant are investigated using 3D hydrodynamical (HD) simulations. Under the assumption that the supernova ejecta is evolved in the medium with a density gradient, the shell is composed of two hemispheres with different radiuses, and the smaller hemisphere is in relatively dense media. The resulting periphery of remnant is consistent with detected ones, and it can be concluded that the peculiar periphery of G349.7+00.2 can be reproduced as the remnants interacting with the medium with a density gradient.
}

   \authorrunning{Yan et al. }            
   \titlerunning{Investigating the morphology of G349.7+00.2}  

   \maketitle

%
%
\section{Introduction}
\label{intro}
SNRs are the candidate of the primary accelerators of the Galactic cosmic rays, and the dynamical evolution and the morphology of them depend heavily on the properties of the ambient medium (e.g., Reynolds~\cite{R08}; Zeng et al. \cite{Zea19}). For example, the SNR G349.7+00.2, which has a distance of $ 22 \mathrm{kpc}$ and the age of $2800 \mathbf{\mathrm{yr}}$ based on the observations of $H_I$ absorption (Caswell et al.~\cite{CM75}), 1720MHz OH masers (Frail et al.~\cite{FG96}) and CO (Reynoso \& Mangum~\cite{RM00}), has a peculiar morphology as indicated in radio and X-rays. The radio image is characterized by a distinct brightness enhancement along the southeastern limb, where the remnant is interacting with a molecular cloud. And the radio structure can be fitted by two overlapping rings, with one smaller and brighter than the other, this feature can be considered that a smaller ring corresponds to a higher density of the inflated shell (Manchester~\cite{MR87}). In X-rays, G349.7+0.2 was first detected in the Galactic plane survey of ASCA (Yamauchi et al.~\cite{YK98}). The analysis of the ASCA data (Slane et al.~\cite{SC02}) showed that the spectrum was well fit by a single-temperature non-equilibrium ionization mode. The Chandra observations also revealed a compact central object inside the SNR shell, called CXOUJ171801.0--372617. The X-ray brightness is enhanced in the southeastern limb, and the overall morphology is similar to that observed in the radio (Lazendic~\cite{LS05}).

The peculiar morphology of G349.7+00.2 shed light on the ambient medium. Reynoso et al.~(\cite{RM00}) proposed that the shock front of the remnant has been extended to the density gradient, and the smaller half of the shell has been expanded to a higher density of gas. Moreover, CO observations toward G349.7+0.2 also revealed a molecular cloud associated with the SNR (Reynoso et al.~\cite{RM00};Reynoso\&Mangum. ~\cite{RM01};Lazendic et al. ~\cite{LW02}). Five OH (1720MH)  maser spots were found in the center of the SNR (Frail et al.~\cite{FG96}), and the shock-excited near-infrared $H_I$ emission has been found toward the center of the remnant as well as OH absorption (1665 and 1667 MHz) in the remnant (Lazendic et al.~\cite{LW10}). The line emission from several molecular transitions shows clear evidence of interaction between the SNR and the molecular clouds (MCs) (Reynoso et al.~\cite{RM00};Lazendic et al.~\cite{LW10};Dubner et al.~\cite{DG04}).

Ergin et al.~(\cite{ES15}) analyzed the GeV $\gamma$-rays from G349.7+0.2 and found that the detected spectrum can be represented by a broken power law distribution of $\alpha=1.89$ and $\beta=2.42$ and a spectral break at $ 12 \,\mathrm{GeV}$.  Zhang et al.~(\cite{ZL16}) constrained the parameters of the escaping-diffusion model using the Markov Chain Monte Carlo method, and they found that the correction factor of slow diffusion around this SNR, i.e.,  $\chi$=0.01 for the power-law injection and $\chi$= 0.1 for the $\delta$-function injection. The $\gamma$-ray spectrum can be fit with a reasonable molecular cloud mass. In addition,  Hnatyk $\&$ Petruk ~(\cite{HP99}) performed 2D hydrodynamical modeling of the evolution and X-ray emission of SNRs expanding in a large-scale density gradient, and the results showed that the remnant expanded into a density gradient.

Three-dimensional(3D) hydrodynamic/ magnetohydrodynamic (HD/ MHD) simulations are widely used to study the morphologies of SNRs (Toledo-Roy ~\cite{TE14}; Fang et al.~\cite{FY17}; Fang et al.~\cite{FY18}; Fang et al.~\cite{FY19}). In this paper, we investigate the reason for the formation of the periphery of G349.7+00.2 using 3D HD simulation under the assumption that the supernova ejecta is evolved in the medium with a density gradient. A detailed description of our model and initial setup of simulations are described in Section 2. The results are presented in Section 3, and some discussion and conclusions are given in Section 4.

\section{Simulation setup}
\label{simuset}
In order to investigate the dynamical evolution of the remnant G349.7+00.2 and reproduce the special morphology, we adopt 3D HD numerical simulations on it with the Pluto code (Mignone et al.~\cite{MB07}; Mignone et al.~\cite{MZ12}), which provides a variety of hydrodynamic modules and algorithms for solving a complete set of MHD/HD equations with different geometry systems. In our simulations, Cartesian geometry (x, y, z) and a computational cube of 6$\times$6$\times$6 $\mathrm{pc}^{-3}$ with $512\times512\times512 $ grids are used.

We assume that the ejecta of G349.7+00.2 has evolved in the nonuniform medium with a density gradient, which directs along $\hat{\xi}=\sin \theta \cos \phi \hat{e}_x + \sin \theta \sin \phi \hat{e}_y + \cos \theta \hat{e}_z$ . After the supernova explosion, the ejecta has a mass of $M_\mathrm{ej} =8.0 M_{\odot}$, a kinetic energy of $E_\mathrm{ej} =10^{51}\mathrm{erg} $ and a radius of $R_\mathrm{ej}=0.5\mathrm{pc}$, which is set at the center of the simulation. The initial matter distribution for the ejecta includes an inner core having a constant density $\rho_c$ within the $r_c$ radius, and an outer layer with a power-law density profile (Fang et al. ~\cite{FY19}), i.e.,
\begin{equation}
\rho_\mathrm{ej}(\mathrm{r})=
\begin{cases}
~\rho_\mathrm{c},&\text {if}~ r<r_\mathrm{c} \\
~\rho_\mathrm{o}(r/R_\mathrm{ej})^\mathrm{-S},&\text{if}~r_\mathrm{c}<r<R_\mathrm{ej}\\
\end{cases}
\end{equation}

Where $\rho_\mathrm{0}$ is the density at $r=R_\mathrm{ej}$.  $\eta$ is the mass ratio of the outer part of the ejecta to the inner one. The index $S=9$ is used in this paper for the core-collapse supernova explosion. The density of the inner core is (Jun \& Norman ~\cite{JN96})
\begin{equation}
\rho_\mathrm{c}=\frac{3(1-\eta)M_\mathrm{ej}}{4\pi r_\mathrm{c}^3},
\end{equation}
Moreover, $r_\mathrm{c}$  can be obtained with
\begin{equation}
r_\mathrm{c}=R_\mathrm{ej}[1-\frac{\eta(3-\mathrm{S})M_\mathrm{ej}}{4\pi \rho_\mathrm{0} R_\mathrm{ej}^3}]^{1/(3-\mathrm{S})},
\end{equation}
The velocity of the matter in the ejecta at $\mathrm{r}$ is $ v=\frac{\mathrm{r}}{R_\mathrm{ej}}v_\mathrm{0}$ for  $r<=R_{\rm ej}$. The velocity of ejecta at the outer edge $v_\mathrm{0}$ can be obtained by
\begin{equation}
 v_\mathrm{0}=(E_\mathrm{ej})^\frac{1}{2}\{\frac{2\pi\rho_\mathrm{c} r_\mathrm{c}^5}{5R_\mathrm{ej}^2}+\frac{2\pi\rho R R_\mathrm{ej}^3[1-(R_\mathrm{ej}/r_\mathrm{c})^{S-5}]}{5-S}\}^{-1/2}.
\end{equation}

\begin{table*}
  \centering
  \caption{Parameters for the modes with different polar angles ($\theta$) and different azimuth angles ($\phi$) with $E_\mathrm{ej} =10^{51}\mathrm{erg}$, $R_\mathrm{ej}=0.5\mathrm{pc}$, $M_\mathrm{ej} =8.0 M_{\odot}$, $n_\mathrm{0}=10\mathrm{cm}^{-3}$, $k=50\mathrm{cm}^{-3}\mathrm{pc}^{-1}$.}\label{2}
  \begin{tabular}{c|cccccc}
     \hline
     Parameter & mode A & mode B & mode C & mode D & mode E & mode F \\
     \hline
     $\theta(^\circ)$ & 60 & 120 & 200 & 120 & 120 & 60 \\
     $\phi(^\circ)$ & 250 & 250 & 250 & 60 & 180 & 300 \\
     \hline
  \end{tabular}
\end{table*}

\begin{table*}
  \centering
  \caption{Parameters for the different modes in the simulations. The common parameters are $E_\mathrm{ej} =10^{51}\mathrm{erg} $, $R_\mathrm{ej}=0.5\mathrm{pc}$, $M_\mathrm{ej} =8.0 M_{\odot}$, $n_\mathrm{0}=10\mathrm{cm}^{-3}$, $\theta=120^{\circ}$ and $\phi=250^{\circ}.$}\label{2}
  \begin{tabular}{c|cccccc}
     \hline
     Parameter & mode H & mode I & mode J & mode K & mode L  \\
     \hline
      ($k\, \mathrm{cm}^{-3} \mathrm{pc}^{-1}$) & 8 & 15 & 25 & 30 & 40  \\
     \hline
  \end{tabular}
\end{table*}

Assuming that the ejection evolved in an environmental medium with a density  gradient, the number density of environmental media satisfies the following formula,
\begin{equation}
n(\bm{r}))=
\begin{cases}
~n_\mathrm{0} + k \hat{\xi}\cdot \bm r,&\text {if}~ \hat{\xi}\cdot \bm r > 0 \\
~n_\mathrm{0},&\text ~otherwise\\
\end{cases}
\end{equation}
where $k$ is the magnitude of the density gradient. $\mu=1.4$ is the mean atomic mass for a gas of a 10 : 1 H:He ratio.

Neglecting the radiative cooling and the particle acceleration involved in the remnant, the dynamical properties of the ejecta can be simulated based on the Euler equations, i.e.
\begin{equation}
\frac{\partial\rho}{\partial t}+\bigtriangledown\cdot(\rho V)=0,
\end{equation}
\begin{equation}
\frac{\partial\rho V}{\partial t}+\bigtriangledown\cdot(\rho VV)+\bigtriangledown P=0,
\end{equation}
\begin{equation}
\frac{\partial E}{\partial t}+\bigtriangledown\cdot(E+P)V)=0,
\end{equation}
and
\begin{equation}
E=\frac{P}{\gamma-1}+\frac{1}{2}\rho V^2,
\end{equation}
where $P$ and $E$ are the gas pressure and the total energy density, respectively. $\gamma=5/3$ is adopted for the nonrelativistic gas and $V$ is the gas velocity.

\section{Results}

G349.7 + 00.2 has a peculiar and interesting morphology in the radio and $X$-rays, and we intend to study the periphery of SNR G349.7 +00.2 through studying the effect of different density gradients. We assume that there is a density gradient in the ambient medium directed along  $\hat{\xi}=\sin \theta \cos \phi \hat{e}_x + \sin \theta \sin \phi \hat{e}_y + \cos \theta \hat{e}_z$. Firstly, $n_\mathrm{0}$ is set to $10\mathrm{cm}^{-3}$ based on the studies in (Tian \& Leahy ~\cite{TL14}). With $M_\mathrm{ej} =8.0 M_{\odot}$, $E_\mathrm{ej} =10^{51}\mathrm{erg} $ can lead to a radius consistent with the observations, so we adopt these values in the simulations. As listed in Table 1, the six modes with different combinations are calculated to seek a reasonable polar angle $\theta$ and an azimuth angle $\phi$ to reproduce the detected periphery of the remnant.

\begin{figure}[tbh]
\begin{center}
\includegraphics[width=0.8\textwidth]{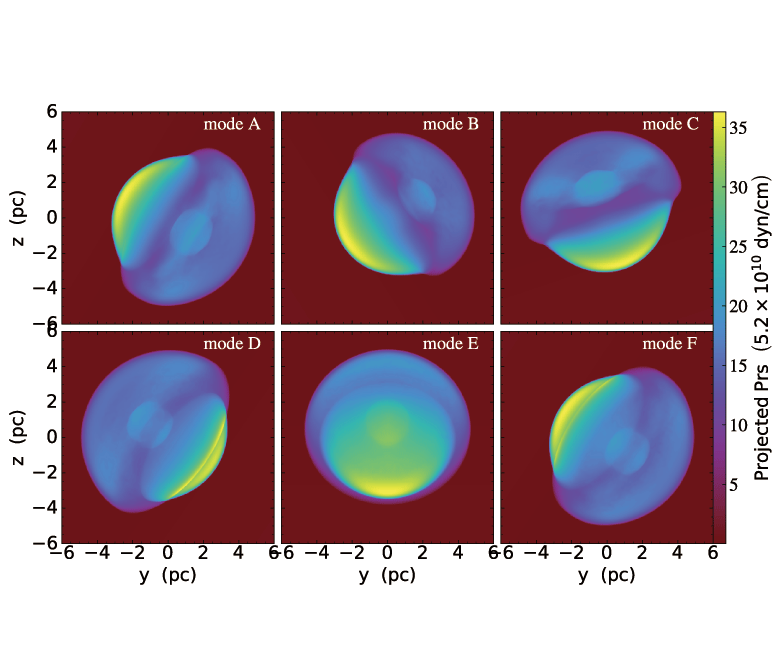}
\end{center}
\caption{The projected pressure along the $x$ direction at t = 2800 yr for the modes A - F with different polar angles ($\theta$) and azimuth angles ($\phi$). The other parameters are $E_\mathrm{ej} =10^{51}\mathrm{erg} $, $R_\mathrm{ej}=0.5\mathrm{pc}$, $M_\mathrm{ej} =8.0 M_{\odot}$, $n_\mathrm{0}=10\mathrm{cm}^{-3}$, $k=50\,\mathrm{cm}^{-3}\mathrm{pc}^{-1}$.}
\label{Fig.ah}
\end{figure}

\begin{figure}[tbh]
\begin{center}
\includegraphics[width=0.5\textwidth]{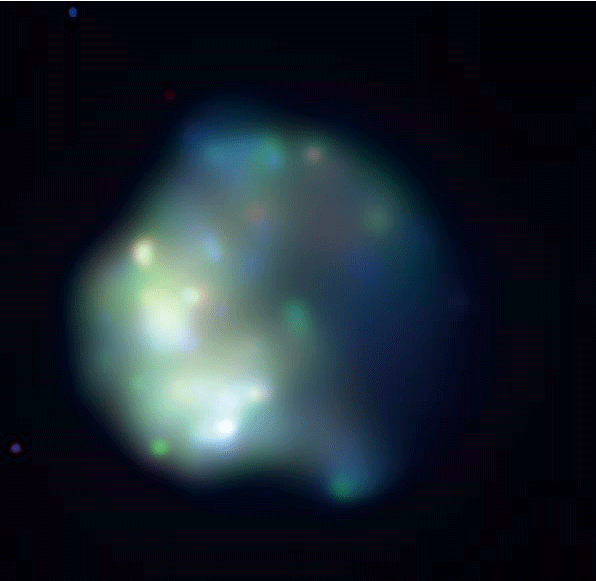}
\end{center}
\caption{The observed X-ray morphology of SNR G349.7+00.2 with Chandra. This image is copied from the website http://hea-www.harvard.edu /ChandraSNR /gallery$\-$gal.html.}
\label{Fig.x-ray}
\end{figure}

Figure \ref{Fig.ah} shows the morphologies of the projected pressure along the $x$ direction at $t= 2800 \mathbf{\mathrm{yr}}$  for these different polar angles $\theta$ and the azimuth angles $\phi$.   With $\theta=120^\circ$ (modes B, D, and E), the morphology is not found to be composed of two hemisphere with an azimuth angle much less than $180^{\circ}$ as illustrated in the mode D. The influence of the polar angle on the resulting morphology of the projected pressure can be seen in the upper three panels (modes A - C) with $\phi = 250^\circ$, and that with $\theta\sim120^\circ$ can be constrained by the image detected by Chandra (Figure \ref{Fig.x-ray}).

\begin{figure}[tbh]
\begin{center}
\includegraphics[width=0.8\textwidth]{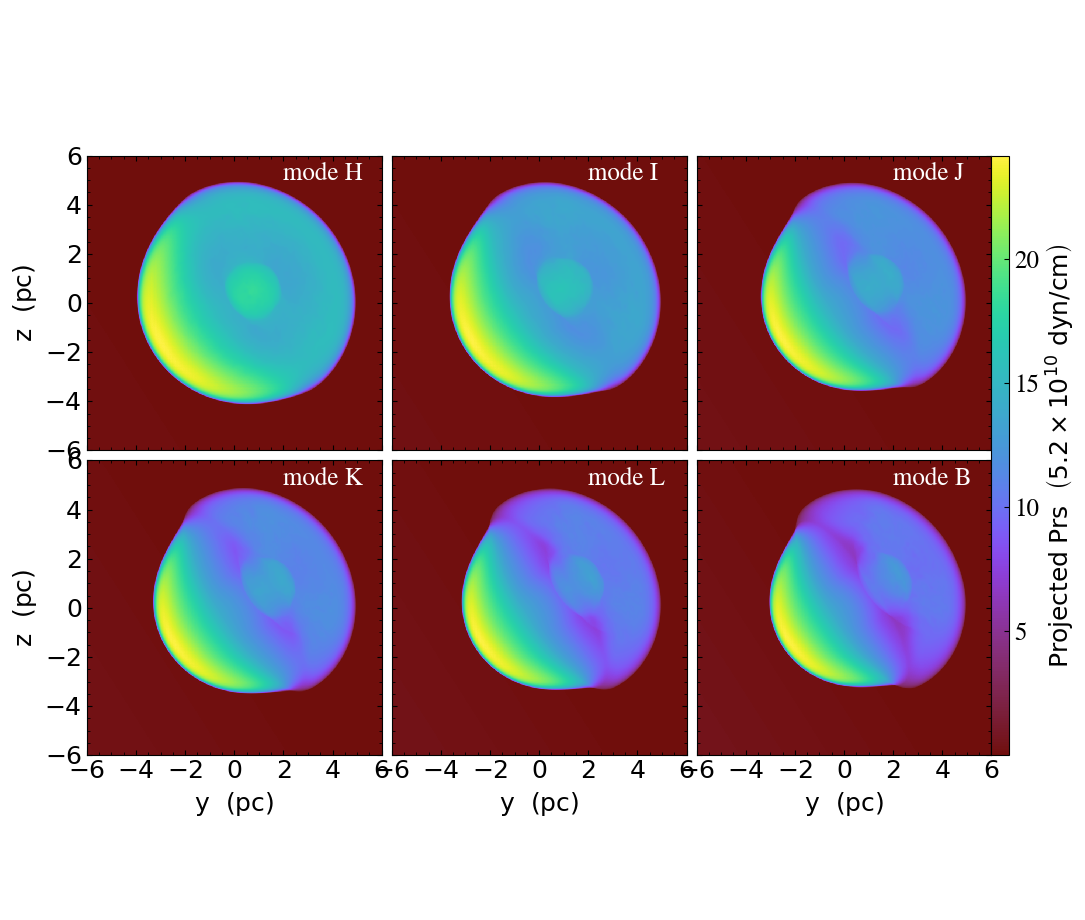}
\end{center}
\caption{The projected pressure along the $x$ direction at $t= 2800 \mathbf{\mathrm{yr}}$ for the modes H - L and mode B. The common parameters are $E_\mathrm{ej} =10^{51}\mathrm{erg} $, $R_\mathrm{ej}=0.5\mathrm{pc}$, $M_\mathrm{ej} =8.0 M_{\odot}$, $n_0=10\mathrm{cm}^{-3}$, $\theta=120^{\circ}$ and $\phi=250^{\circ}$.}
\label{Fig.hm}
\end{figure}

\begin{figure}[tbh]
\begin{center}
\includegraphics[width=0.8\textwidth]{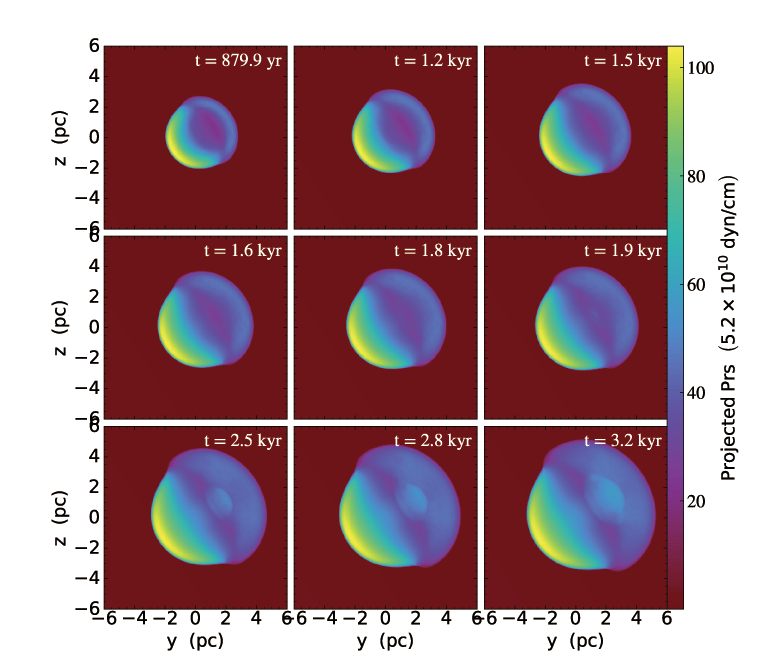}
\end{center}
\caption{The evolution of the projected pressure along the $x$ direction at different times for the mode B. The other parameters are the same as Fig.\ref{Fig.hm}.}
\label{Fig.evo}
\end{figure}

\begin{figure}[tbh]
\begin{center}
\includegraphics[width=0.8\textwidth]{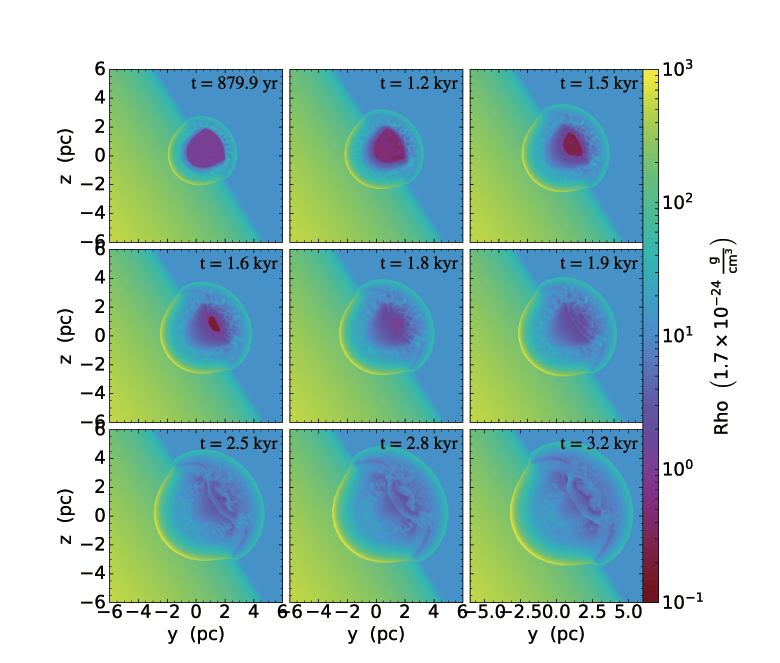}
\end{center}
\caption{The slices of the density in the plane $x=0$ at $t= 2800 \mathbf{\mathrm{yr}}$ for the mode B. The other parameters are the same as Fig.\ref{Fig.hm}.}
\label{Fig.den}
\end{figure}

Figure \ref{Fig.hm} shows the morphologies of the projected pressure along the $x$ direction for the  modes with different density gradient as listed with $\theta = 120^{\circ}$ and $\phi = 250^{\circ}$. The details of the parameters for the modes H - L and B  and are shown in Table 2. All these modes can produce morphologies with two hemispheres. The radius of the hemisphere in the denser medium becomes smaller for a larger density gradient with a  deeper hollow between the two hemispheres. Compared with the detected image with Chandra, the density gradient of $30 - 50 \mathrm{cm}^{-3} \mathrm{pc}^{-1}$ can reproduce a consistent periphery.

Figure \ref{Fig.evo}  shows the morphologies of the projected pressure along the $x$ direction at different times for the mode B, and the slices of density at the plane $x=0$ are indicated in Figure \ref{Fig.den}. Initially, in the two hemispheres, the  forward shock is produced due to the supersonic motion of the ejecta. As the forward shock expands outwardly, the material in the inner part of the remnant becomes tenuous until the thermalized medium drives a reverse shock which continually compresses the matter in the ejecta. {As illustrated in the panel for $t=1.2\,\mathrm{kyr}$ in Figure \ref{Fig.evo}, the reverse shock marks the boundary of the low-density ejecta in the center of the remnant.} Due to the Rayleigh-Taylor instabilities, the fingertip structure is produced near the contact discontinuity.  At a time of $2.5 \mathbf{\mathrm{kyr}}$ , the reverse shocks from the two hemispheres  had encountered near the center of the remnant. On the other hand, the pressure in the southeast (SE) area of the remnant become more significant than the other areas due the higher density. For the mode B, at a time of $t= 2800 \mathbf{\mathrm{yr}}$, the forward shock in the denser medium has a radius of $\sim 3.0 \mathrm{pc}$, whereas it is $\sim 3.4 \mathrm{pc}$ in the other hemisphere, which is consistent with the image obtained with Chandra for a distance of $ 22 \mathrm{kpc}$.

\section{Summary and discussion}
\label{sumdis}

G349.7+00.2 has a peculiar morphology in the $X$-rays. The dynamical evolution and the morphology of a SNR depend heavily on the anisotropy and inhomogeneity of both the ejecta and the ambient medium. In this paper, we investigate the reason for the formation of the peculiar periphery as indicated in the observations for the SNR G349.7+00.2 based on 3D HD simulation. In the model, we assume that the ejecta of G349.7+00.2 has evolved in the ambient medium with a density gradient. The shell consists of two hemispheres with different radius, and the resulting morphology of the projected pressure varies with the different density gradient. Moreover, the peculiar periphery illustrated in the detected images can be reproduced with the model.

With a distance of $ 22 \mathrm{kpc}$, the main shell of  SNR G349.7+00.2 has two different hemispheres with a radius ratio of about $1 : 1.13$ based on the observed image by Chandra, which is consistent with the mode B at the time of  $2800$ yr with $E_\mathrm{ej}=10^{51}\mathrm{erg}$, $M_\mathrm{ej} =8.0 M_{\odot}$ and $n_\mathrm{0}=10\mathrm{cm}^{-3}$.  With $\theta=120^{\circ}$ and $\phi=250^{\circ}$, we find the resulting profile of the remnant in the mode B is similar to that in the $X$-ray image. Therefore, $k=50\,\mathrm{cm}^{-3}\mathrm{pc}^{-1}$ is suitable to explain the special morphology of SNR G349.7+00.2 and our results support the assumption of the mode. The $X$-ray image is characterized by a distinct brightness enhancement along the  southeastern limb, which are illustrated in the image of the projected pressure. Moreover, the shell in the northeast is bright in the X-ray image. Our numerical simulation results show that the bright ridge can be explained as the ejecta interacting with the medium with a higher density.

G349.7+00.2 is one of the radio and $X$-ray brightest SNRs in the Galaxy, and there is much evidence that it is interacting with molecular clouds. Through the analysis of the gamma-ray data, it is considered that G349.7+0.2 is expanding in a dense medium of molecular clouds and emitting gamma-rays by interacting with clumps of molecular material (Ergin et al.~\cite{ES15}).  Previous researches claimed that the surrounding molecular gas caused the peculiar morphologies (Lazendic et al.~\cite{LS05};Castro et al.~\cite{CS10}). Alternatively, Yasumi et al.~(\cite{YN14}) proposed that asymmetric explosions could produce such morphologies. In this paper, we assume that this morphology is due to the evolution of the ejecta in a medium with a density gradient, and use $3$D HD simulation to reproduce its special periphery.

\begin{acknowledgements}
JF is partially supported by the Natural Science Foundation of China (NSFC) through grants 11873042, the Yunnan Applied Basic Research Projects under (2018FY001(-003)), the National Key R$\&$D Program of China under grant No.2018YFA0404204, the Candidate Talents Training Fund of Yunnan Province (2017HB003) and the Program for Excellent Young Talents, Yunnan University (WX069051, 2017YDYQ01).
\end{acknowledgements}

\label{lastpage}

\end{document}